\documentclass[seceq]{ptptex}

\usepackage{graphicx}
\def\bm{\boldsymbol}

\title{
Topological Excitations in Spinor Bose-Einstein Condensates
}

\author{
Yuki \textsc{Kawaguchi}$^1$,
Michikazu \textsc{Kobayashi}$^2$,
Muneto \textsc{Nitta}$^3$,
and 
Masahito \textsc{Ueda}$^{1,4}$
}

\inst{
$^1$Department of Physics, University of Tokyo, \\
Hongo 7-3-1, Bunkyo-ku, Tokyo 113-0033, Japan\\
$^2$Department of Basic Science, Graduate School of Arts and Sciences,\\
University of Tokyo, Komaba 3-8-1, Meguro-ku, Tokyo 153-8902, Japan\\
$^3$Department of Physics, and Research and Education Center for Natural Sciences, \\
Keio University, Hiyoshi 4-1-1, Yokohama, Kanagawa 223-8511, Japan\\
$^4$ERATO Macroscopic Quantum Control Project, JST, Tokyo 113-8656, Japan
}

\abst{
A rich variety of order parameter manifolds of multicomponent
Bose-Einstein condensates (BECs) admit various kinds of topological excitations, such as fractional vortices, monopoles, skyrmions, and knots.
In this paper, we discuss two topological excitations in spinor BECs: 
non-Abelian vortices and knots.
Unlike conventional vortices, non-Abelian vortices neither
reconnect themselves nor pass through each other, but create a rung
between them in a topologically stable manner.
We discuss the collision dynamics of non-Abelian vortices in the cyclic phase of a spin-2 BEC.
In the latter part, 
we show that
a knot, which is a unique topological object characterized by a
linking number or a Hopf invariant [$\pi_3 (S^2)=Z$],  can be created using a conventional quadrupole magnetic field in a
cold atomic system.
}

\begin{document}

\maketitle

\section{Introduction}

Topological excitations exist
in a wide variety of systems, such as liquid crystals, superfluids, superconductors, and our universe.
Among them, Bose-Einstein condensates (BECs) of dilute atomic gases~\cite{Review, Review2}
are the ones of the ideal testing grounds for investigating topological excitations.
The novelty of this system lies in the unprecedented controllability, which enables us to study static and dynamic properties of topological excitations.
Various experimental parameters such as the temperature, atom number, trapping potential,
and even the interatomic interaction can be controlled with high precision.

When a BEC is trapped in a magnetic potential, the spin of each atom is oriented along the direction of the local magnetic field.
The spin degrees of freedom are therefore frozen and the BEC is described by a scalar order parameter.
In such a system, the order parameter manifold is $U(1)$, which can host quantized vortices.
By using high-resolution imaging techniques,
nonlinear dynamical phenomena such as vortex nucleation and vortex-lattice-formation dynamics~\cite{Madison2001}, 
and quantum turbulence~\cite{Henn2009} have been directly observed.

On the other hand, when a BEC is confined in an optical trap, the direction of the spin can change
dynamically due to the interparticle interaction.
Consequently a BEC of spin-$f$ atoms is described with the $(2f+1)$-component order parameter.
In this case, the order parameter manifold is generally larger than $U(1)$, and therefore, various nontrivial topological objects can be accommodated in a spinor BEC.

In this paper, we briefly overview the symmetry properties of the spinor BEC,
and discuss two topological excitations:
non-Abelian vortices (Sec.~\ref{sec:nonAbelian}) and knot solitons (Sec.~\ref{sec:knots}).

\section{Symmetry property of spinor BECs}
\label{sec:spinorBEC}
We consider a system of spin-$f$ atoms under zero magnetic field.
For simplicity, we consider a spatially uniform system.
The Hamiltonian of this system is given by
\begin{align}
 H =& \int d{\bm r}  \frac{\hbar^2}{2M}\sum_m \nabla \hat{\psi}_m^\dagger({\bm r}) \nabla \hat{\psi}_m({\bm r}) \nonumber\\
&+ \int d{\bm r}d{\bm r}'\sum_{mnm'n'}V^{mn}_{m'n'}({\bm r},{\bm r}')\hat{\psi}_m^\dagger({\bm r})\hat{\psi}_n^\dagger({\bm r}') \hat{\psi}_{n'}({\bm r}')\hat{\psi}_{m'}({\bm r}),
\label{eq:hamiltonian}
\end{align}
where $\hat{\psi}_m$ is the field operator for an atom in the magnetic sublevel $m=0,\pm 1,\cdots \pm f$, and $M$ is the atomic mass.
In dilute gases of neutral atoms, the main interatomic interaction is Van der Waals attraction and hard-core repulsion,
which conserves the total spin $\mathcal{F}$ of two colliding atoms.
Each scattering channel of total spin $\mathcal{F}$ can be described with the {\it s}-wave scattering length $a_\mathcal{F}$,
and $V^{mn}_{m'n'}({\bm r},{\bm r}')$ is then given by
\begin{align}
V^{mn}_{m'n'}({\bm r},{\bm r}') = \delta({\bm r}-{\bm r}')
\sum_{\mathcal{F}=0,2,\cdots, 2f} \frac{4\pi \hbar^2}{M}a_\mathcal{F} \sum_{\mathcal{M}=-\mathcal{F}}^\mathcal{F}
\langle fmfn|\mathcal{F}\mathcal{M}\rangle \langle \mathcal{F}\mathcal{M}|fm'fn'\rangle,
\end{align}
where $\langle fmfn|\mathcal{F}\mathcal{M}\rangle$ are the Clebsch-Gordan coefficients.
The scattering amplitude for odd $\mathcal{F}$ vanishes due to the Bose symmetrization.

Since the interaction conserves the total spin and the number of particles,
the Hamiltonian~\eqref{eq:hamiltonian} has the $SO(3)$ rotational symmetry in spin space and the $U(1)$ global gauge symmetry,
i.e., the full symmetry of this system is given by
\begin{align}
G=SO(3)_F \times U(1)_{\phi},
\end{align}
where subscripts $F$ and $\phi$ denote the spin and gauge, respectively.
An element of $G$ can be described with the combination of the Euler rotation $e^{-iF_z\alpha}e^{-iF_y\beta}e^{-iF_z\gamma}$ and
the gauge transformation $e^{i\phi}$,
where $F_\mu$ ($\mu=x,y,z$) is the $\mu$ component of the spin operator and $\alpha, \beta, \gamma$ are the Euler angles.
While the system has the symmetry of $G$ above the transition temperature,
the symmetry $G$ is broken to its subgroup $H$ below the transition temperature.
The subgroup $H$ is referred to as the isotropy group.
The order parameter manifold is defined as the coset space $R=G/H$.

\begin{figure}[ht]
\begin{center}
\includegraphics[width=0.8\linewidth]{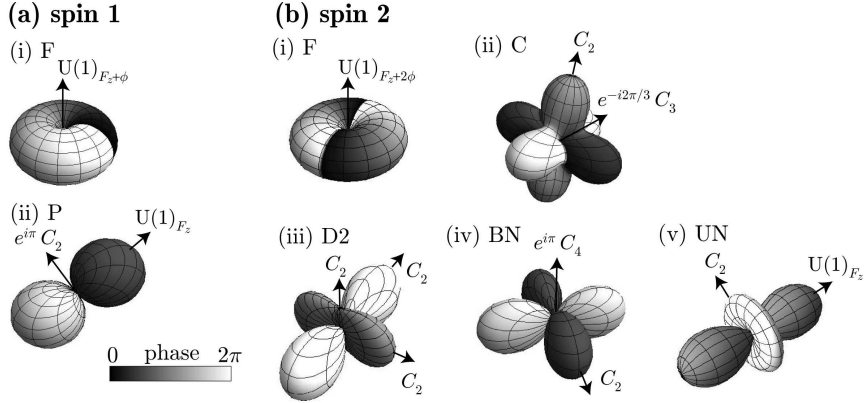}
\end{center}
\caption{
Surface plots of the order parameter defined in Eq.~\eqref{eq:SH}, where the gray scale on the surface represents ${\rm arg}\,\Psi$.
F, P, C, UN, and BN denote ferromagnetic, polar, cyclic, uniaxial nematic, and biaxial nematic phases, respectively.
The arrows labeled with $U(1)$ indicate the continuous symmetry axes, and those labeled with $C_n$ denote the $n$-fold symmetry axis.
For example, in (a) (ii), the order parameter is axisymmetric around the U(1) axis,
and has 2-fold symmetry about the $C_2$ axis;
when we rotate the order parameter about the $C_2$ axis by $\pi$, the phase of the order parameter changes by $\pi$, and
to compensate this phase change, the gauge transformation $e^{i\pi}$ should be applied.
}
\label{fig:OP}
\end{figure}

In the case of spinor BECs, several ground-state phases arise depending on the scattering lengths.
In Table~\ref{table1}, we show
the representing order parameters of the ground state that appear in spin-1 and 2 systems, together with their isotropy groups $H$.
The symmetry of the order parameter can be visualized by plotting the order parameter
\begin{align}
 \Psi(\theta,\phi) = \sum_{m=-f}^f Y_{fm}(\theta,\varphi) \psi_m,
\label{eq:SH}
\end{align}
where $Y_{fm}(\theta,\varphi)$ is the rank-$f$ spherical harmonics with  $(\theta,\varphi)$ denoting the direction in the spin space.
Figure~\ref{fig:OP} shows the surface plots of the ground-state order parameters in spin-1 and spin-2 BECs.
In the case of spin-1 BEC~\cite{Ohmi1998, Ho1998},
the ferromagnetic order parameter has an $SO(2)\cong U(1)$ symmetry axis as shown in Fig.~\ref{fig:OP} (a) (i).
This corresponds to the fact that the order parameter ${\bm \psi}_{\rm F}=(1,0,0)$ 
is invariant under a spin rotation $e^{-iF_z\phi}$ followed by a gauge transformation $e^{i\phi}$,
i.e., the ferromagnetic phase has the spin-gauge coupled $U(1)$ symmetry.
To clarify this particular feature of the coupling between spin and gauge degrees of freedom, we denote this symmetry as $U(1)_{F_z+\phi}$.
On the other hand, the polar state shown in Fig.~\ref{fig:OP} (a) (ii) has $\mathbb{Z}_2$ symmetry in addition to the $U(1)$ symmetry;
${\bm \psi}_{\rm P}=(0,1,0)$ is invariant under $e^{i\pi}e^{-iF_x\pi}$ and $e^{-iF_z\phi}$.
Therefore, the isotropy group $H$ of the polar phase is given by the spin-gauge coupled dihedral group $(D_\infty)_{F_z,\phi}= U(1)_{F_z}\rtimes (\mathbb{Z}_2)_{F,\phi}$,
where semidirect product $\rtimes$ implies that when the nontrivial element of $\mathbb{Z}_2$ acts on an en element $g\in U(1)$, $g$ changes to $g^{-1}$.

The symmetry for the spin-2 order parameters can be investigated in a similar manner, and the results are summarized in Table~\ref{table1}.
There are three mean-field ground states; ferromagnetic, cyclic and D2~\cite{Chibanu2000,Koashi2000},
and quantum fluctuations divide the region of the D2 phase into uniaxial nematic (UN) and biaxial nematic (BN) phases~\cite{UN-BN}.
As shown in Fig.~\ref{fig:OP} (b), the cyclic, D2, UN, and BN phases have discrete symmetries.
In particular, the isotropy groups $H$ for the cyclic, D2, and BN phases are non-Abelian.
In these phases, vortices exhibit nontrivial collision dynamics as discussed in the next section.

\begin{table}
\begin{center}
\begin{tabular}{llllccc}
    & phase & $\bm\psi$             & $H$        & $\pi_1(G/H)$   & $\pi_2(G/H)$ & $\pi_3(G/H)$ \\
\hline\hline
spin-1 & F  & $(1,0,0)$            & $U(1)_{F_z-\phi}$       & $(\mathbb{Z}_2)_{F,\phi}$ & 0            & $(\mathbb{Z})_{F,\phi}$ \\
       & P  & $(0,1,0)$            & $(D_\infty)_{F_z,\phi}$ & $(\mathbb{Z})_{F,\phi}$   & $(\mathbb{Z})_F$ & $(\mathbb{Z})_F$ \\
\hline
spin-2 & F  & $(1,0,0,0,0)$        & $U(1)_{F_z-2\phi}$      & $(\mathbb{Z}_4)_{F,\phi}$ & 0            & $(\mathbb{Z})_{F,\phi}$ \\
       & C  & $(1,0,0,\sqrt{2},0)$ & $(T)_{F,\phi}$          & $(T^*\times \mathbb{Z})_{F,\phi}$          & 0            & $(\mathbb{Z})_F$ \\
       & D2 & $(a,0,b,0,a)$        & $(D_2)_{F_z,\phi}$      & $(Q\times \mathbb{Z})_{F,\phi}$   & 0 & $(\mathbb{Z})_F$ \\
       & UN & $(0,0,1,0,0)$        & $(D_\infty)_{F_z}$      & $(\mathbb{Z}_2)_{F}\times (\mathbb{Z})_\phi$   & $(\mathbb{Z})_F$ & $(\mathbb{Z})_F$ \\
       & BN & $(1,0,0,0,1)$        & $(D_4)_{F_z,\phi}$      & $(D_4^*\times \mathbb{Z})_{F,\phi}$   & 0            & $(\mathbb{Z})_F$ \\
\hline
\end{tabular}
\end{center}
\caption{Representing order parameter $\bm \psi$, isotropy group $H$ of the order parameter, and the first, second and third homotopy groups of $G/H$
for the ground states of spin-1 and 2 BECs.
F, P, C, UN, and BN denote ferromagnetic, polar, cyclic, uniaxial nematic, and biaxial nematic, respectively.
$D_n$ is the dihedral group, $T$ is the tetrahedral group, and $Q$ is the quaternion group.
$T^*$ and $D_4^*$ are defined by $SO(3)/T\cong SU(2)/T^*$ and $SO(3)/D_4\cong SU(2)/D_4^*$, respectively.
}
\label{table1}
\end{table}

The $n$-dimensional topological objects are classified with the $n$th homotopy group $\pi_n(G/H)$,
which are also listed in Table~\ref{table1}.
Due to the rich variety of the order parameter manifolds, various kinds of topological excitations can exist in spinor BECs~\cite{r1,r2,r3,r4,r5,r6,r7,r8,Semenoff2007, Kobayashi2009, Kawaguchi2008}.

\section{Non-Abelian vortices}
\label{sec:nonAbelian}

The topological charge of a vortex determines how the order parameter changes as one circumnavigates a loop surrounding the vortex. 
For example, if the phase of the order parameter of a scaler BEC changes as $e^{in\phi}$, the topological charge of this vortex is said to be $n$. 
In a similar manner, the vortices can be characterized with an operator which is acted on the order parameter as one goes around the vortex.
To satisfy the singlevaluedness the order parameter, the operator should keep the order parameter invariant.

We have shown in Ref.~\citen{Kobayashi2009} that the cyclic phase of a spin-2 BEC can host non-Abelian vortices.
To understand this, we note that the order parameter of the cyclic phase is invariant under the following 12 elements of the tetrahedral group $T$~\cite{Semenoff2007}:
$\Vec{1}$, $I_x = e^{i F_x \pi}$, $I_y = e^{i F_y \pi}$, $I_z = e^{i F_z \pi}$, $\bar{C} = e^{2 \pi i / 3} e^{- 2 \pi i (F_x + F_y + F_z) / 3 \sqrt{3}}$, $\bar{C}^2$, $I_x \bar{C}$, $I_y \bar{C}$, $I_z \bar{C}$, $I_x \bar{C}^2$, $I_y \bar{C}^2$, and $I_z \bar{C}^2$, where we choose lobes in Fig.~\ref{fig:OP} (b) (ii) as $x$, $y$ and $z$ axes.
These operators are regarded as topological charges of vortices in a cyclic BEC,
and therefore, the vortices are non-Abelian.
Furthermore, these operators are classified into four conjugacy classes: 
\begin{itemize}
\item
(I) integer vortices: $\{\Vec{1}\}$;
\item
(II) 1/2 - spin vortices: $\{I_x,I_y,I_z\}$;
\item
(III) 1/3 vortices: $\{\bar{C},I_x \bar{C},I_y \bar{C},I_z \bar{C}\}$;
\item
(IV) 2/3 vortices: $\{\bar{C}^2,I_x \bar{C}^2,I_y \bar{C}^2,I_z \bar{C}^2\}$.
\end{itemize}
By acting an element of $T$, topological charges in the same conjugacy class transform into one another,
which means that the topological charges are not uniquely determined.


\begin{figure}[ht]
\begin{center}
\resizebox{0.9\hsize}{!}{
\includegraphics{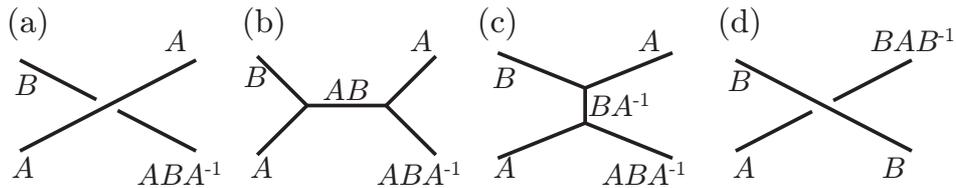}
}
\end{center}
\caption{Collision dynamics of two vortices.
(a) Initial configuration, where 
$A$ and $B$ represent operators that generate the corresponding vortex (a set of spin rotations and gauge transformations for the case of a cyclic BEC).
The vortex on the bottom right, which is connected to $B$, is identified as $ABA^{-1}$.
The configuration in (a) is topologically equivalent to (b) and (c), where a rung is formed.
If the vortices are a pair of vortex and anti-vortex, i.e., $A=B^{-1}$, the rung in (b) disappears, giving rise to reconnection,
whereas the rung in (c) corresponds to a doubly quantized vortex.
If $A$ and $B$ are commutative, passing through is also possible because the configurations of (a) and (d) will then be topologically equivalent.
However, when $A$ and $B$ are not commutative, the collision always results in the formation of a rung.
}
\label{fig:nonAbelian1}
\end{figure}

\begin{figure}[ht]
\begin{center}
\resizebox{0.8\hsize}{!}{
\includegraphics{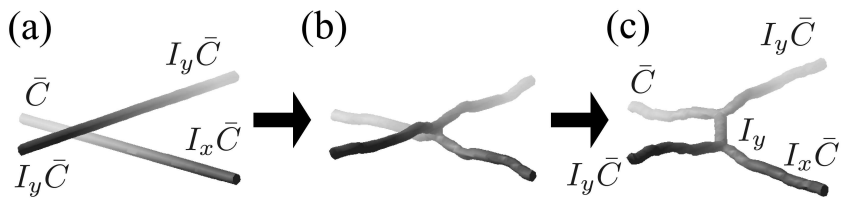}
}
\end{center}
\caption{Numerical simulation for the collision dynamics of non-Abelian vortices in a spin-2 cyclic BEC.
Reprinted from Ref.~\citen{Kobayashi2009}.
}
\label{fig:nonAbelian2}
\end{figure}

The non-Abelian characteristics of the vortices manifest themselves most dramatically in the collision dynamics. In general, when two vortices collide, they reconnect themselves, pass through [Fig.~\ref{fig:nonAbelian1}(d)], or form a rung that bridges the two vortices [Fig.~\ref{fig:nonAbelian1}(b),(c)]. When two Abelian vortices collide, all these three cases are possible, and one of them occurs depending on the kinematic parameters and initial conditions.
However, when two non-Abelian vortices collide, only a rung can be formed,
whereas reconnection and passing through are topologically forbidden because the corresponding operators do not commute with each other.
In fact, the nonvanishing commutator of the two operators gives the operator of the rung vortex.
In Ref.~\citen{Kobayashi2009}, we have numerically simulated the rung formation dynamics
in the spin-2 cyclic phase. 
Figure~\ref{fig:nonAbelian2} illustrates a typical rung formation.

When the core of a 1/3 vortex in the cyclic phase is filled with the ferromagnetic state, which is possible for a certain parameter set,
it is possible to observe such dynamics of vortex lines by using a phase-contrast imaging technique that can detect local magnetization.

\section{Knot soliton}
\label{sec:knots}
Next we consider topological excitations in higher dimensions.
The spin-1 polar phase and spin-2 UN phase can host 't Hooft-Polyakov monopoles and knots,
while other phases in Table~\ref{table1} support three-dimensional skyrmions.
Among them, knots are distinguished from other topological excitations, such as vortices, monopoles, and skyrmions, in that knots are classified by
a linking number while the others are classified by a winding number.
Knots are characterized by mappings from a three-dimensional sphere $S^3$ to $S^2$.
The same topological objects are discussed in high energy physics,
where knots are suggested to exist as stable solitons in a three-dimensional
classical field theory~\cite{Faddeev1997}. 
Here we explain that knots of spin textures can be created and observed in the polar phase of a spin-1 BEC~\cite{Kawaguchi2008}.

\begin{figure}[ht]
\begin{center}
\includegraphics[width=0.9\linewidth]{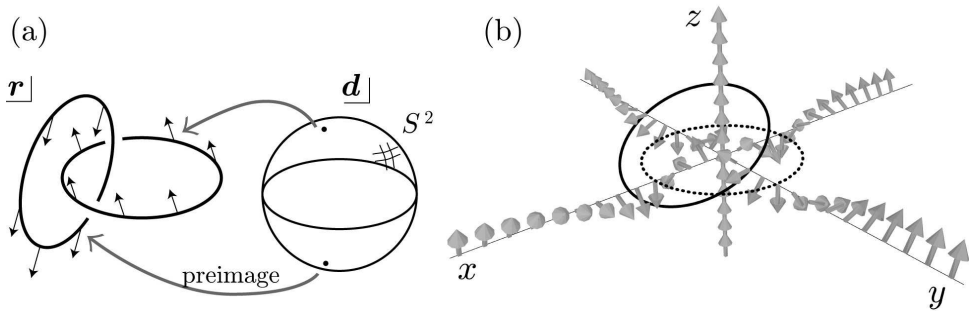}
\end{center}
\caption{
(a) Preimages of two distinct points on $S^2$ forming a link.
(b) Spin configuration of a knot with Hopf charge 1 in a polar BEC,
where the arrows show the ${\bm d}$ field of the polar phase.
The solid and dashed curves trace the point where ${\bm d}$ points to $x$ and $-z$, respectively, forming a link.
Reprinted from Ref.~\citen{Kawaguchi2008}.
}
\label{fig:knot_preimage}
\end{figure}

The $S^3$ domain is prepared by imposing the boundary condition that the order parameter takes on the same value 
in every direction at spatial infinity.
Although the order-parameter manifold for the polar phase is complicated
$R_{\rm P}=(SO(3)_F\times U(1)_\phi)/(D_\infty)_{F,\phi}\cong(S^2_F\times U(1)_\phi)/(\mathbb{Z}_2)_{F,\phi}$,
neither $U(1)$ nor $\mathbb{Z}_2$ symmetry contributes to homotopy groups in spaces higher than one dimension.
Therefore, when we discuss topological objects in higher dimensions,
the order parameter can be described with a unit vector $\bm d \in S^2$ which corresponds to the direction of the ``dumbbell'' in Fig.~\ref{fig:OP} (a) (ii).
Consequently, we have $\pi_3(R_{\rm P})\cong \pi_3(S^2)\cong\mathbb{Z}$.
The associated integer topological charge $Q$ is known as the Hopf charge:
\begin{align}
Q=\frac{1}{4\pi^2} \int d^3x\ \epsilon_{ijk}\mathcal{F}_{ij} \mathcal{A}_k,
\label{eq:H-charge}
\end{align}
where $\mathcal{F}_{ij}=\partial_i\mathcal{A}_j-\partial_j\mathcal{A}_i={\bm d}\cdot(\partial_i {\bm d} \times \partial_j {\bm d})$~\cite{Faddeev1997}.
Note that the domain (${\bm r}$) is three-dimensional, while the target space (${\bm d}$) is two-dimensional.
Consequently, the preimage of a point on target $S^2$ constitutes a closed loop in $S^3$. Furthermore, the Hopf charge is interpreted as the linking number of these loops: 
if the ${\bm d}$ field has Hopf charge $Q$,
two loops corresponding to the preimages of any two distinct points on the target $S^2$ will be linked $Q$ times [see Fig.~\ref{fig:knot_preimage} (a)].
Figure~\ref{fig:knot_preimage} (b) shows an example of the ${\bm d}$ field of a polar BEC with Hopf charge 1%
\footnote{Strictly speaking, the configuration in Fig.~\ref{fig:knot_preimage} is an {\it unknot},
since the preimage of one point on $S^2$ forms a simple ring which is unknotted.}.

\begin{figure}[ht]
\begin{center}
\includegraphics[width=0.7\linewidth]{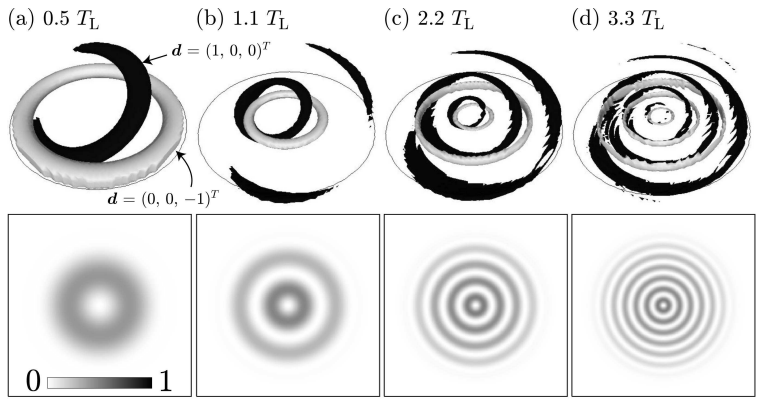}
\end{center}
\caption{Dynamics of the creation of knots in a spherical optical trap under a quadrupole magnetic field.
Snapshots of the preimages of ${\bm d}=(0,0,-1)^{\rm T}$ and ${\bm d}=(1,0,0)^{\rm T}$ (top), as well as
cross sections of the density for $m=-1$ components on the $xy$ plane (bottom).
Reprinted from Ref.~\citen{Kawaguchi2008}.
}
\label{fig:knottime}
\end{figure}

Knots can be created by manipulating an external magnetic field.
In the presence of an external magnetic field,
the linear Zeeman effect causes the Larmor precession of ${\bm d}$,
while ${\bm d}$ tends to become parallel to the magnetic field because of the quadratic Zeeman effect.
Suppose that we prepare an optically trapped BEC in the $m=0$ state [i.e., ${\bm d}=(0,0,1)^{\rm T}$] 
by applying a uniform magnetic field in the $z$ direction.
Then, we suddenly turn off the uniform field and switch on a quadrupole field.
Because of the linear Zeeman effect, ${\bm d}$ starts rotating around the local magnetic field,
and therefore the ${\bm d}$ field winds as a function of $t$, resulting in a formation of knots.
Figure~\ref{fig:knottime} shows the creation dynamics of knots in an optical trap subject to the quadrupole field,
where the upper panels show the snapshots of the preimages of ${\bm d}=-\hat{z}$ and ${\bm d}=\hat{x}$
and the lower panels show cross sections of the density for $m=-1$ components on the $xy$ plane.
The density pattern in $m=-1$ components is the smoking gun of the knots;
a double-ring pattern appears that corresponds to one knot.
As the ${\bm d}$ field winds in the dynamics, the number of rings increases.
This prediction can be tested by the Stern-Gerlach experiment.

\section{Concluding remarks}
\label{sec:conclusion}
We have discussed the symmetry properties of spinor Bose-Einstein condensates (BECs) and topological excitations in them.
In particular, we have shown that when the BEC has the symmetry of non-Abelian group, the collision dynamics of the non-Abelian vortices are quite different from
those of conventional vortices in a scalar BEC and superfluid $^4$He.
We have also shown that the spin-1 polar phase and spin-2 uniaxial nematic phase can accommodate a knot.
The knot is unique in that
topological excitations are classified with a linking number, while other topological excitations are classified with a winding number.
The superfluid helium-3 is often referred to as a testing ground to simulate our universe, 
because it can accommodate various topological excitations similar to those known in gauge-field theory.
The variety of topological excitations in spinor BECs is as rich as that in superfluid helium-3.
The great advantage of spinor BECs is that direct observation and manipulation of the topological excitations
are possible, while complicated analysis is required when using superfluid helium-3 if one is to identify a topological excitation from NMR signals.
The controllability of spinor BECs
would stimulate our imagination, thereby promoting further developments in the study of topological excitations in the superfluid systems.


%

\end{document}